\documentclass[12pt]{iopart}
\usepackage{amssymb}
\usepackage{amsfonts}
\usepackage{psfrag}
\usepackage{graphicx}
\usepackage{color}
\usepackage{cite}
\eqnobysec

\def\keywords#1{\vspace{10pt}
     \begin{indented}
     \item[]\rm Keywords: #1\par
     \end{indented}}

\def\be{\begin{equation}}
\def\ee{\end{equation}}
\def\bea{\begin{eqnarray}}
\def\eea{\end{eqnarray}}

\def\CL{{\mathcal L}}
\def\CR{{\mathcal R}}
\def\CT{{\mathcal T}}
\def\CS{{\mathcal S}}
\def\CN{{\mathcal N}}

\def\fT{{\mathfrak T}}

\def\fS{{\mathfrak S}}
\def\fP{{\mathfrak P}}

\def\scal{\stackrel{\rm s.l.}{=}}

\begin{document}
\jl{1}

\title[Reaction-diffusion on the fully-connected lattice]{Reaction-diffusion on the fully-connected lattice:
\boldmath{$A+A\rightarrow A$}}

\author{Lo\"\i c Turban and Jean-Yves Fortin}

\address{Groupe de Physique Statistique, Laboratoire de Physique et Chimie Th\'eoriques, Universit\'e de Lorraine--CNRS, 
Vand\oe uvre l\`es Nancy Cedex, F-54506, France} 

\ead{loic.turban@univ-lorraine.fr, jean-yves.fortin@univ-lorraine.fr}

\begin{abstract}
Diffusion-coagulation can be simply described by a dynamic where particles perform
a random walk on a lattice and coalesce with probability unity when meeting on the same site.
Such processes display non-equilibrium properties with strong 
fluctuations in low dimensions. 
In this work we study this problem on the fully-connected lattice, an 
infinite-dimensional system in the thermodynamic limit, for which mean-field behaviour is expected. 
Exact expressions for the 
particle density distribution at a given time and survival time distribution for a given number of particles 
are obtained. In particular we show that the time needed to reach a 
finite number of surviving particles (vanishing density in the scaling limit) displays strong fluctuations 
and extreme value statistics, characterized by a universal class of 
non-Gaussian distributions with singular behaviour. 
\end{abstract}


\submitto{J. Phys. A: Mathematical and Theoretical}

\section{Introduction} 
Cooperative phenomena in strongly interacting systems out of equilibrium
are often difficult to describe with standard approximations which usually 
fail. 
This is especially true in low dimensions where fluctuations are essential and relevant. 
Simple models often help to shed light on the physical processes at work
and give access to statistical distributions that are important for the 
universal classification of these models~\cite{hinrichsen00,odor04,henkel08,odor08,krapivsky10}.

The study of the kinetics of irreversible reaction-diffusion processes have been the subject of much 
interest during the last forty years. From the theoretical point of view these processes can be sufficiently 
simple to offer the possibility of exact 
solutions~\cite{ovchinnikov78,bramson80,toussaint83,kang84a,kang84b,peliti86,lushnikov87,spouge88,
bramson88,doering89,privman92,alcaraz94,lee94,lee95,bray03} and their asymptotic behaviour
can be classified among different universality classes~\cite{henkel95a,schutz01,hinrichsen00,odor04,tauber05}. 
They are important in nature as 
well as for applications~\cite{frielander77,zeldovich87,ovchinnikov89,grindrod91}.

The most simple examples are the single-species processes $A+A\rightarrow$ \O\   (diffusion-annihilation) 
and $A+A\rightarrow A$ (diffusion-coalescence) which belong to the same universality class. 
In both cases, above the critical dimension $D_c=2$ the density fluctuations can be neglected and the particle density 
$x$ evolves according to the mean-field rate equation 
\be
\frac{dx}{dt}=-kx^2\,,
\label{rate-equa}
\ee
where $k$ is the reaction-rate constant. The solution for an initial density $x(0)=1$ is then
\be
x(t)=\frac{1}{1+kt}\,.
\label{xt-mf}
\ee
The $t^{-1}$ long-time decay is actually obtained on the Bethe lattice~\cite{ben-avraham07}. 
It is corrected by a logarithmic factor at $D_c$ where $x(t)\simeq(\ln t)/t$~\cite{bramson80,kang84b,peliti86,lee94}.
Below $D_c$, where the density fluctuations are relevant, the kinetic exponent becomes $D$-dependent, 
$x(t)\simeq t^{-D/2}$, as shown by scaling arguments, 
numerical simulations and exact results~\cite{toussaint83,kang84a,kang84b,peliti86,lushnikov87,spouge88,lee94}.  
A $t^{-1/2}$ decay has been indeed observed experimentally in effectively one-dimensional
systems~\cite{prasad89,kopelman90,kroon93}.

In the present work we study the statistical properties of the diffusion-coalescence 
process ($A$\ +\ \O\ $\rightarrow$\ \O\ +\ $A$
and $A+A\rightarrow A$) on the fully-connected lattice (complete graph) with $N$ sites, the absorbing state
consisting of one particle left. In the limit $N\to\infty$ such a lattice 
can only be embedded in an infinite-dimensional space, thus one expects a mean-field behaviour. 
Our main motivation is to obtain, besides exact results for mean values, exact expressions for different
probability distributions, in particular for extreme values. This is a continuation of 
previous works on stochastic processes on the fully-connected lattice by one of us~\cite{turban14,turban15}.

Our main results can be summarized as follows in the scaling limit (s.l.).
The mean number $s$ of particles surviving at time $t$ and its variance behave as:
\be
\frac{\overline{s_N(t)}}{N}\scal\frac{1}{t+1}\,,\qquad
\frac{\overline{\Delta s_N^2(t)}}{N}\scal\kappa(t)=\frac{1}{3(t+1)}\left[1-\frac{3t+1}{(t+1)^3}\right]\,.
\label{sNtDsN2t0}
\ee
The probability density $\fS(\sigma,t)$ associated with the scaled variable
\be
\sigma=\frac{s-\overline{s_N(t)}}{N^{1/2}}\,,
\label{sigmat0}
\ee
is a Gaussian with variance $\kappa(t)$ given in~\eref{sNtDsN2t0}.

Let $t$ be the time needed to reach a number $v$ of surviving particles (first-passage time through $v$).
Its statistical properties depend on the value of $v$. When $v=O(N)$ one obtains
\be
\overline{t_N(x)}\scal\frac{1}{x}-1\,,\qquad 
N\overline{\Delta t_N^2(x)}\scal\chi(x)=\frac{2}{3}-\frac{1}{x}+\frac{1}{3x^3}\,.
\label{tNxDtN2x0}
\ee
where $x=v/N$.
The fluctuations of the scaled variable
\be
\theta=N^{1/2}\left[t-\overline{t_N(x)}\right]
\label{thetax0}
\ee
are also Gaussian, with variance $\chi(x)$ given by~\eref{tNxDtN2x0}. 

When $x\to0$, i.e. when $v=O(1)$, $\chi(x)$ diverges.
This is the signal of a different scaling behaviour. As a function of $v$, the mean first-passage time 
and its variance are now given by
\be
\frac{\overline{t_N(v)}}{N}\scal\frac{1}{v}\,,\qquad
\frac{\overline{\Delta t_N^2(v)}}{N^2}\scal2\left(\frac{\pi^2}{6}-H_{v,2}\right)+\frac{1}{v}\left(\frac{1}{v}-2\right)\,,
\label{tNvDtN2v0}
\ee
where $H_{v,2}$ is a generalized harmonic number. Since the variance is growing as 
$N^2$ the scaled time variable can be defined as:
\be
\theta'=\frac{t}{N}\,,\qquad\overline{\theta'}=\frac{1}{v}\,,\qquad\theta'\geq0\,.
\label{theta'0}
\ee
The associated probability density then reads
\bea
\fl\fT'(v,\theta')&=\frac{(-1)^v}{v!(v\!-\!1)!}\sum_{l=0}^\infty(-1)^l(2l+1)
\frac{(l+v)!}{(l-v)!}\,\e^{-l(l+1)\theta'}\,,\nonumber\\
\fl&=\frac{1}{v!(v\!-\!1)!}\prod_{m=0}^{v-1}\left[\frac{\partial}{\partial\theta'}+m(m+1)\right]
\frac{1}{2q^{1/4}}\left.\frac{\partial\vartheta_1(z,q)}{\partial z}\right|_{z=0}\,,\qquad q=\e^{-\theta'}\,.
\label{T'vtheta'-7-0}
\eea
where $\vartheta_1(z,q)$ is a Jacobi theta function.
The probability density $\fT'(v,\theta')$ behaves asymptotically as:
\be
\fT'(v,\theta')\simeq\left\{
\begin{array}{ll}
\frac{(2v+1)!}{v!(v\!-\!1)!}\,\exp\left[-v(v+1)\theta'\right]\,,&\qquad \theta'\gg1\,,\nonumber\\
\ms
\frac{1}{2^{2v}v!(v\!-\!1)!}\left(\frac{\pi}{\theta'}\right)^{2v+3/2}
\exp\left(-\frac{\pi^2}{4\theta'}\right)\,,&\qquad\theta'\ll1\,.
\end{array}
\right.
\label{T'vtheta'-9-0}
\ee
It decays exponentially when $\theta'\gg1$ and vanishes with an essential singularity
at $\theta'=0$.

The outline of the paper is the following. In section 2, we study the statistics of the number of 
particles surviving after $k$ updates. After defining the model, the probability distribution 
is obtained by solving the eigenvalue problem associated with the master equation. 
The mean value and the variance are then deduced from a generating function. The section ends 
with a solution of the partial differential equation following from the master equation in 
the scaling limit. Section 3 is concerned with the statistical properties of the first-passage time through 
a given number $v$ of surviving particles. A generating function for the probability distribution 
is first obtained, from which the mean value and the variance are deduced. For $v=O(N)$, 
as above, the probability density is obtained in the scaling limit
by solving a partial differential equation. For $v=O(1)$, where the fluctuations are much stronger 
and non-Gaussian, the scaling limit is deduced directly from the probability distribution. 
The discussion in section 4 is followed by six appendices where some calculational details are given.

\section{Number of particles $\bi s$ surviving after $\bi k$ updates}

\subsection{Model and master equation for $S_N(s,k)$}

We consider a fully-connected lattice with $N$ sites. Initially all the sites are singly occupied.
The system evolves in time $t$ via random sequential updates. 
Let $s$ be the number of particles surviving after $k-1$ updates. During the $k$th update a site $i$ is picked 
at random among the $N$. This site is occupied with probability $s/N$ or empty with probability $1-s/N$. 
When the site is occupied, the selected particle jumps at random on one of the $N$ sites. If the destination 
site $j$ is occupied by another particle, the two particles coalesce. This occurs with probability $(s-1)/N$.
Otherwise, nothing happens. These rules avoid a multiple occupation of the sites. The time $t$ 
is incremented by $1/N$ for each update. 

The evolution of the system during an update can be summarized as follows:
\bea
\fl s(k)&=s(k-1)-1\qquad &\mathrm{with\ probability}\qquad \underbrace{\frac{s}{N}}_{i\ \mathrm{occupied}}\times
\underbrace{\frac{s-1}{N}}_{j\neq i\ \mathrm{occupied}}\,,\nonumber\\
\fl s(k)&=s(k-1)\qquad &\mathrm{with\ probability}\qquad \underbrace{1-\frac{s}{N}}_{i\ \mathrm{empty}}\ 
+\underbrace{\frac{s}{N}}_{i\ \mathrm{occupied}}\times\underbrace{1-\frac{s-1}{N}}_{j\ \mathrm{empty\ or\ }j=i}\,.
\label{model}
\eea
Thus the probability distribution $S_N(s,k)$ for the number $s$ of surviving particles after $k$ updates is 
governed by the following master equation:
\be
S_N(s,k)=\left[1-\frac{s(s-1)}{N^2}\right]S_N(s,k-1)+\frac{s(s+1)}{N^2}S_N(s+1,k-1)\,.
\label{master-1}
\ee

\subsection{Eigenvalue problem}
Introducing a column state vector $|S_N(k)\rangle$ with components $S_N(s,k)$, $s=1,\ldots,N$, 
the master equation~\eref{master-1} can be rewritten in matrix form as
$|S_N(k)\rangle=\mathsf{T}\,|S_N(k-1)\rangle$ where $\mathsf{T}$ is the transition matrix of the Markov chain given by:
\be
\mathsf{T}=\left(\begin{array}{cccccc}
1       &\frac{1\times2}{N^2}  &0	            &0                   &0                 &0                   \\
0       &1\!-\!\frac{1\times2}{N^2}&\frac{2\times3}{N^2} &0                   &0                 &0                   \\
	&                      &\ddots              &\ddots              &                  &                    \\
0	&0                     &0                   &1\!-\!\frac{s(s-1)}{N^2}&\frac{s(s+1)}{N^2}&0                   \\
        &                      &                    &                    &\ddots            &\ddots              \\ 
0	&0	               &0	            & 0	                 &0                 &1\!-\!\frac{N(N-1)}{N^2}\\
\end{array}\right)\,.
\label{tmatrix}
\ee
The eigenvalue problem $\mathsf{T}|v^{(n)}\rangle=\lambda_n|v^{(n)}\rangle$ leads to the following system of equations
\be
\left[1-\frac{s(s-1)}{N^2}-\lambda_n\right]v_s^{(n)}+\frac{s(s+1)}{N^2}v_{s+1}^{(n)}=0\,,\qquad s=1,\ldots,N\,,
\label{eigen-1}
\ee
with $v_{N+1}^{(n)}=0$. The eigenvalues are given by:
\be
\lambda_n=1-\frac{n(n-1)}{N^2}\,,\qquad n=1,\ldots,N\,.
\ee
The corresponding eigenvectors satisfy the following relations:
\bea
v_s^{(n)}&=(-1)^{n-s}\frac{s(s+1)^2(s+2)^2\cdots(n-1)^2n}{(n-s)!(n+s-1)(n+s)\cdots(2n-3)(2n-2)}\,v_n^{(n)}\nonumber\\
\ms
&=(-1)^{n-s}\,\frac{2n-1}{s}{n+s-2\choose 2s-2}{2s-2\choose s-1}{2n-1\choose n}^{-1}v_n^{(n)}\,,\quad s\leq n\,,\nonumber\\
v_s^{(n)}&=0\,,\qquad s>n\,.\nonumber\\
\label{eigen-3}
\eea 
The $v_n^{(n)}$ are left undetermined and depend on the initial state.

\subsection{Probability distribution}
\begin{figure}[!th]
\begin{center}
\includegraphics[width=8cm,angle=0]{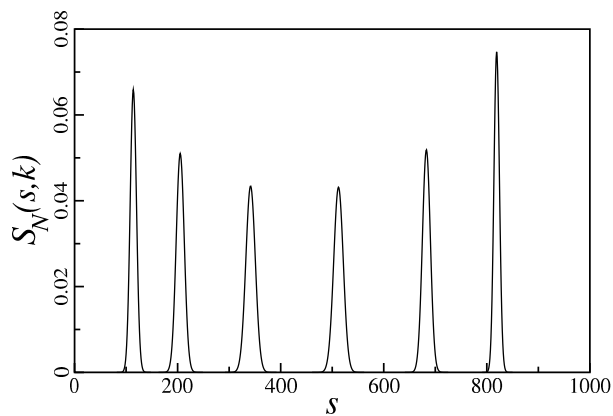}
\end{center}
\vglue -.5cm
\caption{Evolution of the probability distribution $S_N(s,k)$ of the number of particles $s$ remaining after
$k$ updates, obtained through a numerically exact solution of the master equation~\eref{master-1}. 
The initial number of particles is $s=N=1024$ and the number of updates is $k=2^n$ with $n=8$ to 13
from right to left.
\label{fig-1}
}
\end{figure}

We assume that all the sites are occupied by one particle in the initial state so that $S_N(s,0)=\delta_{s,N}$.
Writing the initial state vector as 
\be
|S_N(0)\rangle=\sum_{n=1}^N|v^{(n)}\rangle
\label{sv0}
\ee
and using \eref{eigen-3} one obtains
\be\fl
S_N(s,0)=\sum_{n=s}^Nv_s^{(n)}=\sum_{n=s}^N(-1)^{n-s}\,\frac{2n-1}{s}
{n+s-2\choose 2s-2}{2s-2\choose s-1}{2n-1\choose n}^{-1}v_n^{(n)}\,,
\label{SN0}
\ee
and, as shown in appendix~A, the initial condition is satisfied when
\be
v_n^{(n)}={2n-1\choose n}\prod_{j=0}^{n-1}\frac{N-j}{N+j}\,.
\label{vnn}
\ee
The probability distribution at later time follows from 
\be
|S_N(k)\rangle=\mathsf{T}^k|S_N(0)\rangle=\sum_{n=1}^N\lambda_n^k|v^{(n)}\rangle,
\label{SNk}
\ee
so that finally:
\be\fl
S_N(s,k)=\sum_{n=s}^N(-1)^{n-s}\,\frac{2n-1}{s}
{n+s-2\choose 2s-2}{2s-2\choose s-1}\prod_{j=0}^{n-1}\frac{N-j}{N+j}\left[1-\frac{n(n-1)}{N^2}\right]^k\,.
\label{SNsk}
\ee
The time evolution of $S_N(s,k)$ is illustrated in figure~\ref{fig-1}.

\subsection{Mean value and mean square value}
The mean values can be deduced from the generating function
\be
\CS_N(y,k)=\sum_{s=1}^Nsy^sS_N(s,k)\,,
\label{SNyk-2}
\ee
studied in appendix~A.
According to~\eref{SNyk-1} and~\eref{Sigmany}, the mean value of $s$ after $k$ updates is given by:
\be\fl
\overline{s_N(k)}=\sum_{s=1}^NsS_N(s,k)=\CS_N(1,k)=\sum_{n=1}^N(2n-1)\prod_{j=0}^{n-1}\frac{N-j}{N+j}
\left[1-\frac{n(n-1)}{N^2}\right]^k\,.
\label{sNk}
\ee
The derivative of the generating function $\CS_N(y,k)$ at $y=1$ leads to:
\bea
\fl\overline{s_N^2(k)}&=\sum_{s=1}^Ns^2S_N(s,k)=\left.\frac{\partial\CS_N}{\partial y}\right|_{y=1}\nonumber\\
\fl&=\sum_{n=1}^N(2n-1)\prod_{j=0}^{n-1}\frac{N-j}{N+j}
\left[1-\frac{n(n-1)}{N^2}\right]^k\left[1+\left.\frac{dP_{n-1}(2y-1)}{dy}\right|_{y=1}\right]\,.
\label{sN2k-1}
\eea
The generating function for Legendre polynomials
\be
\sum_{l=0}^\infty P_l(x)r^l=\frac{1}{\sqrt{1-2xr+r^2}}
\label{gfPl}
\ee
can be used to give:
\be
\sum_{l=0}^\infty\left.\frac{dP_l(x)}{dx}\right|_{x=1}r^l=\frac{r}{(1-r)^3}\,,
=\sum_{j=0}^\infty{j+2\choose 2}r^{j+1}\,.
\label{gfdPl}
\ee
Identifying the coefficients of $r^{n-1}$, one obtains:
\be
\left.\frac{dP_{n-1}(2y-1)}{dy}\right|_{y=1}\!\!\!=2\,{n\choose 2}=n(n-1)\,.
\label{dPny1}
\ee
Thus, the mean square value of $s$ after $k$ updates is given by:
\be
\overline{s_N^2(k)}=\sum_{n=1}^N(2n-1)\left[1+n(n-1)\right]\prod_{j=0}^{n-1}\frac{N-j}{N+j}
\left[1-\frac{n(n-1)}{N^2}\right]^k\,.
\label{sN2k-2}
\ee

\subsection{Mean value, variance and probability density in the scaling limit}

\begin{figure}[!th]
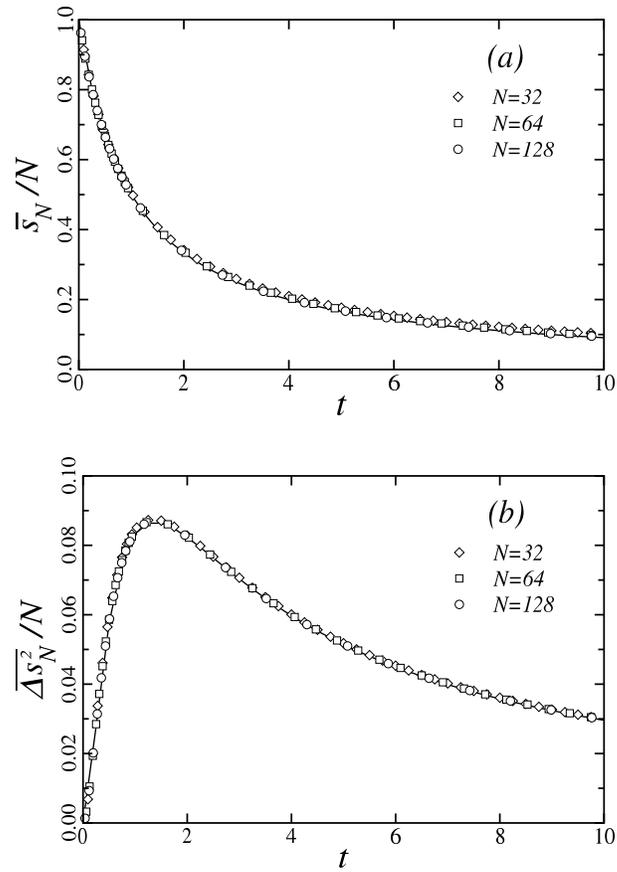

\begin{center}
\includegraphics[width=8cm,angle=0]{fig-2a.eps}
\vglue 5mm
\includegraphics[width=8cm,angle=0]{fig-2b.eps}
\end{center}
\vglue -.5cm
\caption{Scaling behaviour of (a) the mean value $\overline{s_N}$ and (b) the variance $\overline{\Delta s_N^2}$ 
of the number of surviving particles as a function of time $t=k/N$. The data collapse on the full lines 
corresponding to the scaling functions in~\eref{sNtDsN2t}.
\label{fig-2}
}
\end{figure}

\begin{figure}[!th]
\begin{center}
\includegraphics[width=8cm,angle=0]{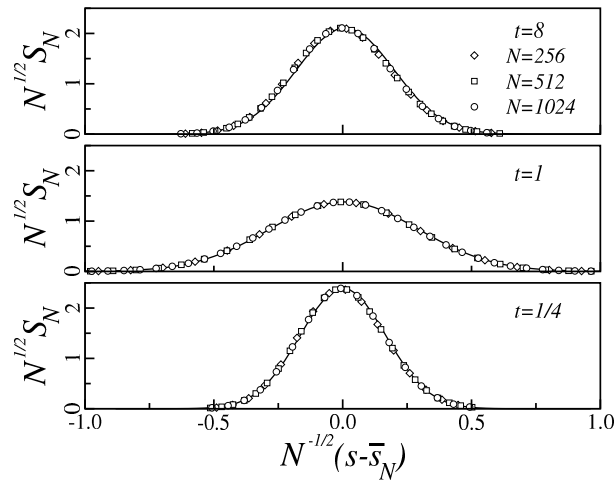}
\end{center}
\vglue -.5cm
\caption{Data collapse for the scaled probability distribution $N^{1/2}S_N(s,k)$ as 
a function of $\sigma=N^{-1/2}(s-\overline{s_N(t)})$ at different times $t=k/N$ 
and for increasing lattice sizes, $N=256$ (diamond), 512 (square) and 1024 (circle). 
The full lines correspond to the Gaussian density~\eref{fP-2} obtained in the scaling limit.
\label{fig-3}
}
\end{figure}

The leading and next-to-leading contributions to the mean value $\overline{s_N(k)}$
and the mean square value $\overline{s_N^2(k)}$ are calculated for $N\gg1$ and used to evaluate the variance in appendix B.

The scaling limit (s.l.) corresponds to $N\to\infty$, $k\to\infty$, $s\to\infty$ for fixed values of $k/N$ and $s/N$.
According to~\eref{sNt} and~\eref{DsN2t}, in this limit one obtains~\footnote[1]{The mean 
value is in agreement with the mean-field result in~\eref{xt-mf}}: 
\be
\frac{\overline{s_N(t)}}{N}\scal\frac{1}{t+1}\,,\qquad
\frac{\overline{\Delta s_N^2(t)}}{N}\scal\kappa(t)=\frac{1}{3(t+1)}\left[1-\frac{3t+1}{(t+1)^3}\right]\,.
\label{sNtDsN2t}
\ee
The time evolution of $\overline{s_N}/N$ and $\overline{\Delta s_N^2}/N$ is shown in figure~\ref{fig-2}.
The behaviour of the variance suggests the introduction of the scaled variables~\footnote[2]{Let $z$ be a discrete random variable in a system of size $N$ with probability distribution $P_N(z)$, mean value     $\overline{z_N}$ and variance  $\overline{\Delta z_N^2}\sim N^{2\alpha}$. The deviation from the mean, $z-\overline{z_N}$, typically grows with $N$ like the standard deviation, i.e., as $N^\alpha$. Thus the ratio $\zeta=(z-\overline{z_N})/N^\alpha$ is a scale-invariant variable. In the scaling limit, the probability density 
$\fP(\zeta)$ is such that $P_N(z)dz\scal\fP(\zeta)d\zeta$ or 
$N^\alpha P_N(z)\scal\fP(\zeta)$.}
\be
\sigma=\frac{s-\overline{s_N(t)}}{N^{1/2}}\,,\qquad t=\frac{k}{N}\,.
\label{sigmat}
\ee
The normalized probability density, $\fS(\sigma,t)$, corresponds to $N^{1/2}S_N(s,k)$. It is obtained
as the solution of the master equation~\eref{master-1} which, in the continuum limit, leads to
\be
\frac{\partial\fS}{\partial t}=\frac{2}{t+1}\left(\fS+\sigma\frac{\partial\fS}{\partial\sigma}\right)
+\frac{1}{2(t+1)^2}\left[1-\frac{1}{(t+1)^2}\right]\frac{\partial^2\fS}{\partial\sigma^2}\,,
\label{dSdt}
\ee
as shown in appendix~C. Now, considering that $\fS(\sigma,t)$ only depends on $t$ through
$\kappa(t)$ defined in~\eref{sNtDsN2t}, equation~\eref{dSdt} can be rewritten as:
\be
\frac{d\kappa}{dt}\left(\frac{\partial\fS}{\partial\kappa}-\frac{1}{2}\frac{\partial^2\fS}{\partial\sigma^2}\right)
=\frac{2}{t+1}\left(\fS+\sigma\frac{\partial\fS}{\partial\sigma}+\kappa\frac{\partial^2\fS}{\partial\sigma^2}\right)\,.
\label{dPdtheta}
\ee
This last equation admits for solution the Gaussian density: 
\be
\fS(\sigma,t)=\frac{\e^{-\sigma^2/[2\kappa(t)]}}{\sqrt{2\pi\kappa(t)}}\,,\qquad
\kappa(t)=\frac{1}{3(t+1)}\left[1-\frac{3t+1}{(t+1)^3}\right]\,,
\label{fP-2}
\ee
Indeed, the Gaussian density is a solution of the diffusion equation on the left and the 
bracket on the right vanishes. Furthermore, it satisfies the initial condition since $\kappa\to0$, $\sigma\to(s-N)/N^{1/2}$
and $\fS(\sigma,\kappa)\to\delta(\sigma)$ as $t\to0$.

The data collapse on the Gaussian density for large $N$ values is shown in figure~\ref{fig-3}.

\section{First-passage time through a number $\bi v$ of surviving particles}
In this section we look for the probability distribution $T_N(v,k)$ of the number $k$ of updates needed to reach a state with $v$
surviving particles. Then $t=k/N$ is the first-passage time through this value $v$. 

\subsection{Generating function}

\begin{figure}[!th]
\begin{center}
\includegraphics[width=10cm,angle=0]{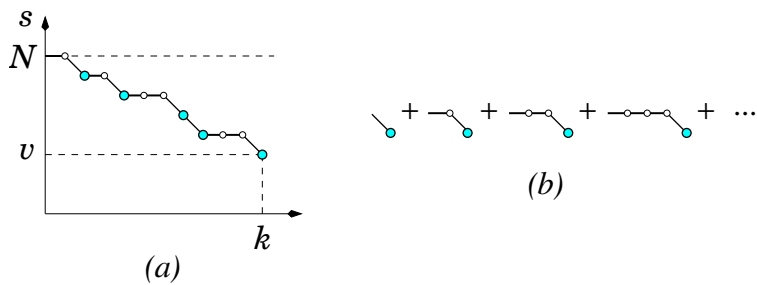}
\end{center}
\vglue -.5cm
\caption{A time evolution of the number of surviving particles is sketched in (a). A small circle corresponds 
to an update where the number of particles stays the same and a bigger circle to an update where two particles
coalesce. Then $t=k/N$ gives the first-passage time through the number of particles $s=v$. 
The diagrams in (b) give the contributions to the generating function \eref{TNvz-2} for the lifetime 
of a sequence of updates ending with a coalescence process. 
\label{fig-4}
}
\end{figure}

The generating function for $T_N(v,k)$ is defined as 
\be
\CT_N(v,z)=\sum_{k=1}^\infty z^kT_N(v,k)\,.
\label{TNvz-1}
\ee
As shown in figure~\ref{fig-4}(a), starting with $s=N$ particles the evolution of the system can be decomposed into a succession 
of steps where the number of particles $s$ remains the same for some time until two particles coalesce. 
Let us associate with these steps 
a generating function for their lifetimes measured in the number of updates $\CL_N(s,z)$. This generating function corresponds to 
the sum of the diagrams shown in figure~\ref{fig-4}(b) and reads:
\bea
\fl\CL_N(s,z)&=\Bigg\{1+z\left[1-\frac{s(s-1)}{N^2}\right]+\cdots
+\underbrace{z^l\left[1-\frac{s(s-1)}{N^2}\right]^l}_{l\ \mathrm{updates\ without\ coalescence}}
+\cdots\Bigg\}\underbrace{z\frac{s(s-1)}{N^2}}_{\mathrm{coalescence}}\nonumber\\
\fl&=\frac{zs(s-1)}{N^2-z\left[N^2-s(s-1)\right]}\,.
\eea
In the sum the coefficient of $z^{l+1}$ corresponds to the probability to have $l$ updates for which $s$ 
remains constant (small circles) 
followed by one update where two particles coalesce (big circles) and $l$ goes from zero to infinity. The generating 
function for $T_N(v,k)$ is obtained as the product of the generating functions for 
the lifetimes with $s$ going from $N$ to $v+1$:
\be
\CT_N(v,z)\!=\!\prod_{s=N}^{v+1}\CL_N(s,z)\!=\!\frac{v}{N}\left(\frac{N!}{v!}\right)^2
\!\!\!\frac{z^{N-v}}{\prod_{s=v+1}^N\left\{N^2-z\left[N^2-s(s-1)\right]\right\}}\,.
\label{TNvz-2}
\ee
It follows from~\eref{TNvz-1} and~\eref{TNvz-2} that $\sum_{k=1}^\infty T_N(v,k)=\CT_N(v,1)=1$ so that the probability 
distribution $T_N(v,k)$ is properly normalized.

\subsection{mean value, mean square value and variance}

\begin{figure}[!th]
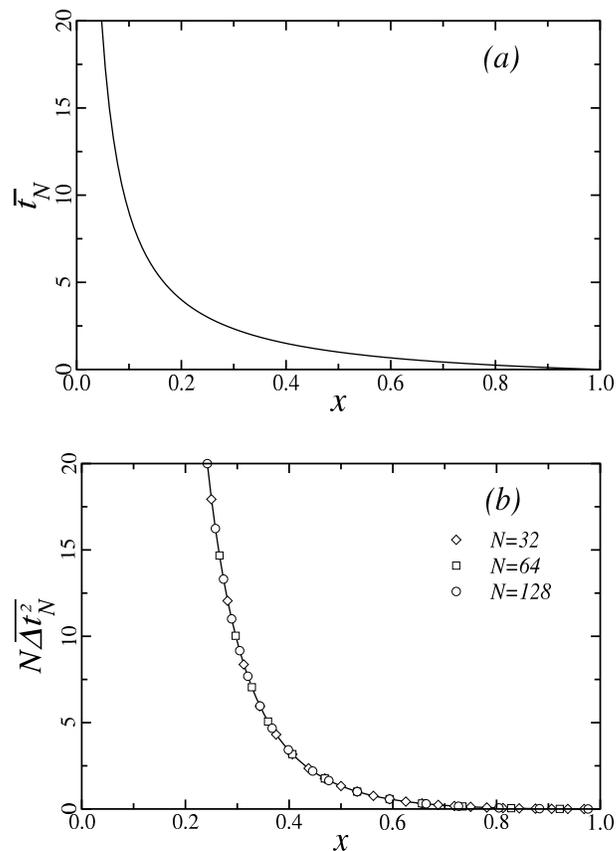

\begin{center}
\includegraphics[width=8cm,angle=0]{fig-5a.eps}
\vglue 5mm
\includegraphics[width=8cm,angle=0]{fig-5b.eps}
\end{center}
\vglue -.5cm
\caption{Behaviour as a function of the particle density $x=v/N$ of (a) the mean value $\overline{t_N}$ and (b) the scaled 
variance $N\overline{\Delta t_N^2}$ of the first-passage time through a given value of $v$. There are no finite-size corrections in (a) and a
good data collapse on the full line 
corresponding to the scaling limit in~\eref{tNxDtN2x} is obtained in (b).
\label{fig-5}
}
\end{figure}

According to~\eref{TNvz-1}, the mean value of the first passage time $t=k/N$ through the value $v$ is given by:
\be
\overline{t_N(v)}=\frac{1}{N}\left.\frac{\partial\CT_N}{\partial z}\right|_{z=1}\,.
\label{tNv-1}
\ee
Making use of
\be
\frac{\partial\CT_N}{\partial z}=\CT_N\frac{\partial\ln\CT_N}{\partial z}
=N^2\frac{\CT_N}{z}\sum_{s=v+1}^N\frac{1}{N^2-z\left[N^2-s(s-1)\right]}\,,
\label{parTNvz}
\ee
one obtains:
\be
\overline{t_N(v)}=N\sum_{s=v+1}^N\frac{1}{s(s-1)}=\frac{N}{v}-1\,.
\label{tNv-2}
\ee
In the same way, we have:
\be
\overline{t_N^2(v)}=\frac{1}{N^2}\left.\frac{\partial}{\partial z}
\left[z\frac{\partial\CT_N}{\partial z}\right]\right|_{z=1}\,.
\label{tN2v-1}
\ee
Some straightforward calculations lead to
\bea\fl
&\frac{\partial}{\partial z}\left[z\frac{\partial\CT_N}{\partial z}\right]
=N^2\frac{\CT_N}{z}\left\{N^2\left[\left(\sum_{s=v+1}^N\!\!\CR_N(s,z)\right)^2\!
+\!\sum_{s=v+1}^N\!\!\CR_N^2(s,z)\right]\!-\!\sum_{s=v+1}^N\!\!\CR_N(s,z)\right\}\nonumber\\
\ms
\fl&\CR_N(s,z)=\frac{1}{N^2-z\left[N^2-s(s-1)\right]}\,,
\label{par2TNvz}
\eea
so that, according to~\eref{tNv-2} and~\eref{tN2v-1}:
\be
\overline{t_N^2(v)}=\overline{t_N(v)}^2+N^2\sum_{s=v+1}^N\left(\frac{1}{s-1}
-\frac{1}{s}\right)^2-\left(\frac{1}{v}-\frac{1}{N}\right)\,.
\label{tN2v-2}
\ee
This can be rewritten as
\be\fl
\overline{t_N^2(v)}=\overline{t_N(v)}^2+2N^2\left(H_{N,2}-H_{v,2}\right)+
\left(\frac{1}{v}-\frac{1}{N}\right)\left[N^2\left(\frac{1}{v}+\frac{1}{N}-2\right)-1\right]\,,
\label{tN2v-3}
\ee
where $H_{n,m}=\sum_{j=1,n}j^{-m}$ is a generalized harmonic number.
Finally the variance is given by:
\be
\overline{\Delta t_N^2(v)}=2N^2\left(H_{N,2}-H_{v,2}\right)+
\left(\frac{1}{v}-\frac{1}{N}\right)\left[N^2\left(\frac{1}{v}+\frac{1}{N}-2\right)-1\right]\,.
\label{DtN2v}
\ee

\subsection{Mean value, variance and probability density in the scaling limit when $v={\mathrm O}(N)$}

When $v={\mathrm O}(N)$ the scaling limit corresponds to $N\to\infty$, $k\to\infty$ and $v\to\infty$
for fixed values of $t=k/N$ and $x=v/N$. Then, according to~\eref{tNv-2}, \eref{DtN2v} and appendix~D, one has:
\be
\overline{t_N(x)}\scal\frac{1}{x}-1\,,\qquad 
N\overline{\Delta t_N^2(x)}\scal\chi(x)=\frac{2}{3}-\frac{1}{x}+\frac{1}{3x^3}\,.
\label{tNxDtN2x}
\ee
The evolution of $\overline{t_N}$ and $N\overline{\Delta t_N^2}$ with the particle density $x$ 
is shown in figure~\ref{fig-5}. The behaviour of the variance leads to the following form of the scaled variables:
\be
\theta=N^{1/2}\left[t-\overline{t_N(x)}\right]=\frac{k}{N^{1/2}}-N^{1/2}\left(\frac{1}{x}-1\right)\,,\qquad x=\frac{v}{N}\,.
\label{thetax}
\ee
The normalized probability density, $\fT(x,\theta)$, is given by the scaling limit of $N^{1/2}T_N(v,t)$.
 
The probability $T_N(v,k)$ that the number of surviving particles reaches the value $v$ after $k$ updates is given by
\be
T_N(v,k)=S_N(v+1,k-1)\frac{v(v+1)}{N^2}\,,
\label{TNvk-1}
\ee
i.e., the probability $S_N(v+1,k-1)$ to have $v+1$ particles after $k-1$ updates multiplied by the probabity 
$v(v+1)/N^2$ of a coalescence process at the next update. With the initial condition $T_N(v,1)=\delta_{v,N-1}$,
\eref{SNsk} and~\eref{TNvk-1} lead to:
\be\fl
T_N(v,k)\!=\!\frac{1}{v!(v\!-\!1)!N^2}\!\sum_{n=v+1}^N\!\!(-1)^{n-v-1}(2n\!-\!1)
\frac{(n\!+\!v\!-\!1)!}{(n\!-\!v\!-\!1)!}\prod_{j=0}^{n-1}\frac{N\!-\!j}{N\!+\!j}\left[1\!-\!\frac{n(n\!-\!1)}{N^2}\right]^{k-1}\!\!\!\!\!\!\!\!\!.
\label{TNvk-2}
\ee
This probability distribution satisfies the master equation
\be
T_N(v,k)=\left[1-\frac{v(v\!+\!1)}{N^2}\right]T_N(v,k\!-\!1)+\frac{v(v+1)}{N^2}\,T_N(v\!+\!1,k\!-\!1)\,,
\label{master-3}
\ee
which follows from~\eref{master-1} and~\eref{TNvk-1}. In the continuum limit, as shown in appendix~E, the following partial differential
equation is obtained:
\be
\frac{\partial\fT}{\partial x}=\frac{1}{2}\left(\frac{1}{x^2}-\frac{1}{x^4}\right)\frac{\partial^2\fT}{\partial\theta^2}\,.
\label{dTdx}
\ee
Assuming that $\fT(x,\theta)=\fT[\chi(x),\theta]$ with $\chi(x)$ given by~\eref{tNxDtN2x} leads to the diffusion equation:
\be
\frac{\partial\fT}{\partial\chi}=\frac{1}{2}\frac{\partial^2\fT}{\partial\theta^2}\,.
\label{dQdchi}
\ee
\begin{figure}[!t]
\begin{center}
\includegraphics[width=8cm,angle=0]{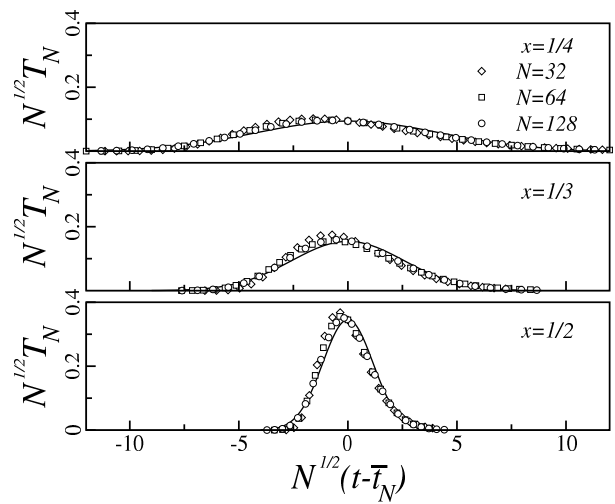}
\end{center}
\vglue -.5cm
\caption{Data collapse for the scaled probability distribution $N^{1/2}T_N(v,k)$ as 
a function of $\theta=N^{1/2}(t-\overline{t_N(x)})$ for different values of the particle density $x=v/N$ 
and for increasing lattice sizes, $N=32$ (diamond), 64 (square) and 128 (circle). 
The full lines correspond to the Gaussian density~\eref{Qchitheta} obtained in the scaling limit.
\label{fig-6}
}
\end{figure}
The Gaussian density
\be
\fT(x,\theta)=\frac{\e^{-\theta^2/[2\chi(x)]}}{\sqrt{2\pi\chi(x)}}\,,\qquad\chi(x)=\frac{2}{3}-\frac{1}{x}+\frac{1}{3x^3}\,.
\label{Qchitheta}
\ee
satisfies the initial condition since  $\chi\to0$, $\theta\to N^{1/2}t$
and $\fT(x,\theta)\to\delta(\theta)$ as $x\to1$. The data collapse on this Gaussian density is shown in figure~\ref{fig-6}.

\subsection{Mean value, variance and probability density in the scaling limit when $v={\mathrm O}(1)$}
\begin{figure}[!th]
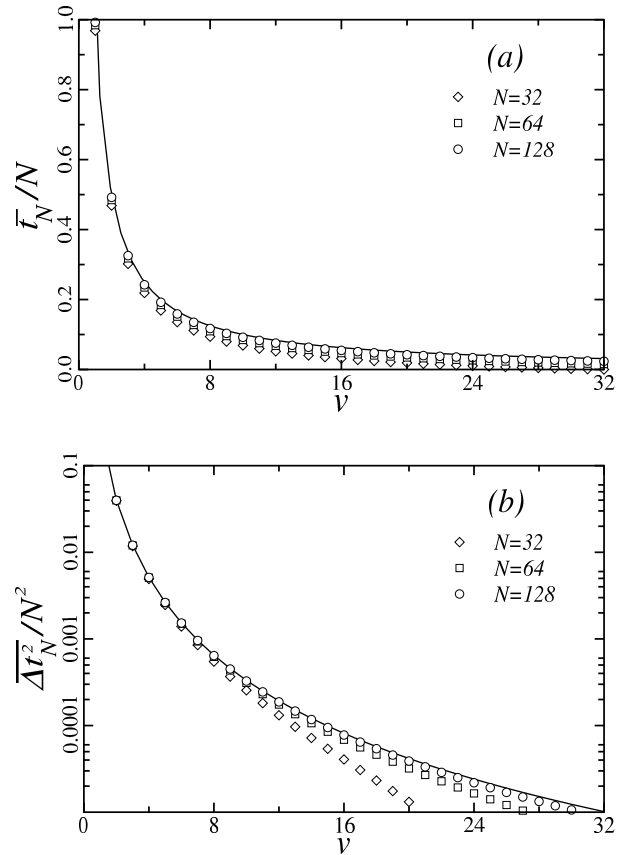

\begin{center}
\includegraphics[width=8cm,angle=0]{fig-7a.eps}
\vglue 5mm
\includegraphics[width=8cm,angle=0]{fig-7b.eps}
\end{center}
\vglue -.5cm
\caption{Scaling behaviour of (a) the mean first-passage time 
$\overline{t_N}$ through a given value of $v$ and (b) its variance
$\overline{\Delta t_N^2}$. With increasing $N$ the data approach the scaling limits in~\eref{tNvDtN2v} (full lines).
\label{fig-7}
}
\end{figure}
\begin{figure}[!t]
\begin{center}
\includegraphics[width=8cm,angle=0]{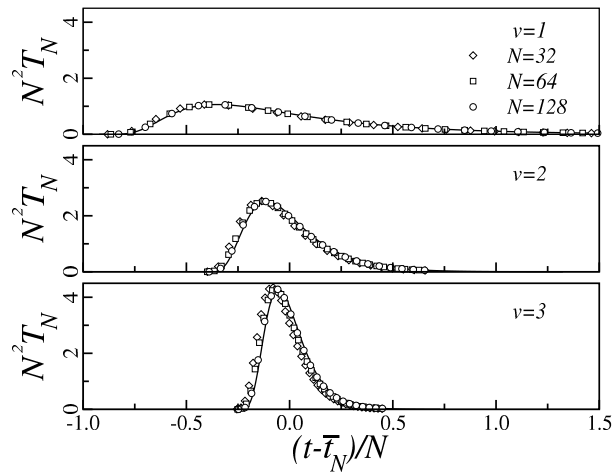}
\end{center}
\vglue -.5cm
\caption{Data collapse for the scaled probability distribution $N^2T_N(v,k)$ as 
a function of $\theta'-1/v=(t-\overline{t_N(v)})/N$ for $v=O(1)$ 
and for increasing lattice sizes, $N=32$ (diamond), 64 (square) and 128 (circle). 
The full lines correspond to the probability density~$\fT'(v,\theta')$ in \eref{T'vtheta'-2}, obtained in the scaling limit.
\label{fig-8}
}
\end{figure}

When $v={\mathrm O}(1)$, according to~\eref{tNv-2} and~\eref{DtN2v} the mean value and the variance of 
the first-passage time through $v$ behave as 
\be
\frac{\overline{t_N(v)}}{N}\scal\frac{1}{v}\,,\qquad
\frac{\overline{\Delta t_N^2(v)}}{N^2}\scal2\left(\frac{\pi^2}{6}-H_{v,2}\right)+\frac{1}{v}\left(\frac{1}{v}-2\right)\,,
\label{tNvDtN2v}
\ee
in the scaling limit (see figure~\ref{fig-7}). The fluctuations are much stronger with a variance now growing as $N^2$ for $\overline{\Delta t_N^2(v)}$ instead of decreasing as $1/N$ for $\overline{\Delta t_N^2(x)}$. 
The scaled time variable is defined as~\footnote[3]{It will be more convenient here to leave 
out the shift by the mean value, $\overline{\theta'}=1/v$.}:
\be
\theta'=\frac{t}{N}=\frac{k}{N^2}\,,\qquad \theta'\geq0\,.
\label{theta'}
\ee
Accordingly, in this limit the probability density is given by:
\be
\fT'(v,\theta')\scal N^2T_N(v,k) 
\label{T'vtheta'-1}
\ee
In the expression~\eref{TNvk-2} of $T_N(v,k)$ as $N\to\infty$ one has
\be
\prod_{j=0}^{n-1}\frac{N\!-\!j}{N\!+\!j}\simeq \e^{-n(n-1)/N}\scal1\,\qquad
\left[1\!-\!\frac{n(n\!-\!1)}{N^2}\right]^{k-1}\scal\e^{-n(n-1)\theta'}\,,
\label{simp}
\ee
for any finite value of $n$, so that the probability density takes the following form:
\be
\fT'(v,\theta')=\!\frac{1}{v!(v\!-\!1)!}\!\sum_{n=v+1}^\infty\!\!(-1)^{n-v-1}(2n\!-\!1)
\frac{(n\!+\!v\!-\!1)!}{(n\!-\!v\!-\!1)!}\,\e^{-n(n-1)\theta'}\,.
\label{T'vtheta'-2}
\ee
The normalization of this probability density is verified in~appendix~F. 
The data collapse on $\fT'(v,\theta')$ is shown in figure~\ref{fig-8}.

The leading contribution for large values of $\theta'\gg1$ is given by the first term in the sum:
\be
\fT'(v,\theta')\simeq\!\frac{(2v+1)!}{v!(v\!-\!1)!}\,\e^{-v(v+1)\theta'}\,,\qquad \theta'\gg1\,.
\label{T'vtheta'-4}
\ee

In order to study the behaviour of $\fT'(v,\theta')$ when $\theta'\ll1$  we first 
show that it can be rewritten 
in terms of derivatives of the Jacobi theta function 
\be
\vartheta_1(z,q)=2q^{1/4}\sum_{l=0}^\infty(-1)^lq^{l(l+1)}\sin[(2l+1)z]\,.
\label{theta1}
\ee
Let us start from~\eref{T'vtheta'-2} with the change $l=n-1$ in the sum, then:
\be
\fT'(v,\theta')=\frac{(-1)^v}{v!(v\!-\!1)!}\sum_{l=v}^\infty(-1)^l(2l+1)
\frac{(l+v)!}{(l-v)!}\,\e^{-l(l+1)\theta'}\,.
\label{T'vtheta'-5}
\ee
Since the ratio of factorials in the sum vanishes when $l=0,\ldots,v-1$ one may write:
\be
\fT'(v,\theta')=\frac{(-1)^v}{v!(v\!-\!1)!}\sum_{l=0}^\infty(-1)^l(2l+1)
\frac{(l+v)!}{(l-v)!}\,\e^{-l(l+1)\theta'}\,.
\label{T'vtheta'-6}
\ee
Then, grouping the extreme factors in pairs, the ratio of factorials can be rewritten as:
\be\fl
\frac{(l+v)!}{(l-v)!}=\prod_{j=1}^v(l-v+j)(l+v-j+1)=\prod_{m=0}^{v-1}[l(l+1)-m(m+1)]\,,\qquad m=v-j\,.
\label{ratio-1}
\ee
Since $l(l+1)$ results from the action of the operator $-\partial/\partial\theta'$ on the exponential term,
one may rewrite~\eref{T'vtheta'-6} as
\bea
\fl\fT'(v,\theta')&=\frac{1}{v!(v\!-\!1)!}\prod_{m=0}^{v-1}\left[\frac{\partial}{\partial\theta'}
+m(m+1)\right]\sum_{l=0}^\infty(-1)^l(2l+1)\,\e^{-l(l+1)\theta'}\,.\nonumber\\
\fl&=\frac{1}{v!(v\!-\!1)!}\prod_{m=0}^{v-1}\left[\frac{\partial}{\partial\theta'}+m(m+1)\right]
\frac{1}{2q^{1/4}}\left.\frac{\partial\vartheta_1(z,q)}{\partial z}\right|_{z=0}\,,\qquad q=\e^{-\theta'}\,,
\label{T'vtheta'-7}
\eea
where $\vartheta_1(z,q)$ is the Jacobi theta function defined in~\eref{theta1}.

\begin{figure}[!t]
\begin{center}
\psfrag{Y}[Bc][Bc][1][1]{$\mathfrak{T}'(v,\theta')$}
\psfrag{X}[tc][tc][1][0]{$\theta'$}
\psfrag{a}[Bc][Bc][1][1]{\tiny $v=1$}
\psfrag{b}[Bc][Bc][1][1]{\tiny $v=2$}   
\psfrag{c}[Bc][Bc][1][1]{\tiny $v=3$}   
\includegraphics[width=8cm,angle=0]{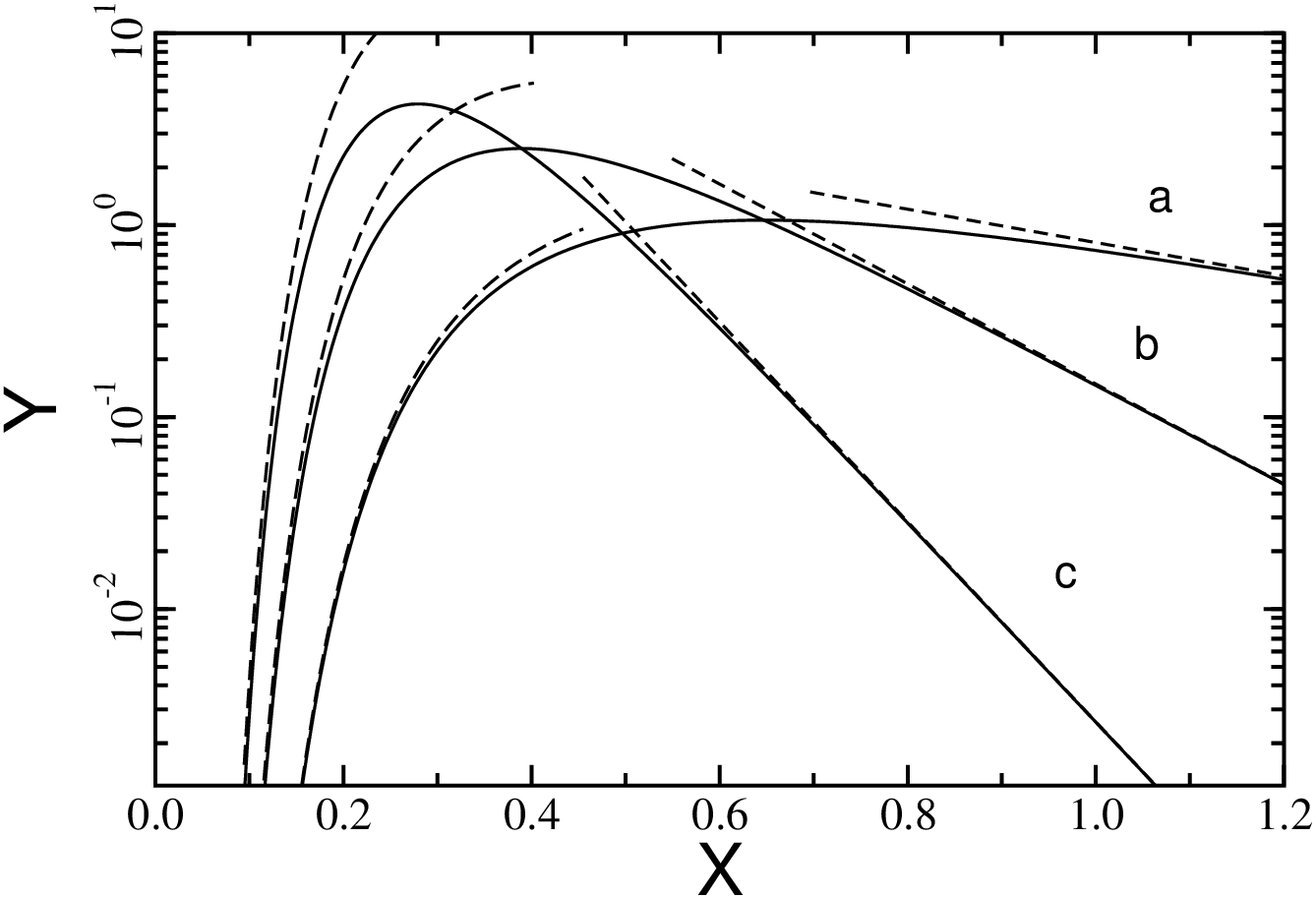}
\end{center}
\vglue -.5cm
\caption{Probability density~$\fT'(v,\theta')$ 
(full lines) and its asymptotic behaviour \eref{T'vtheta'-4} when $\theta'\gg1$
and \eref{T'vtheta'-9} when $\theta'\ll1$ (dashed lines)
\label{fig-9}
}
\end{figure}

The behaviour in the short scaled-time limit, $\theta'\ll1$, can be obtained 
using Jacobi's imaginary transformation~\cite{whittaker27}. Introducing the notation
\be
\vartheta_1(z|\tau)=\vartheta_1(z,q)\,,\qquad q=\e^{i\pi\tau}\,,\qquad \tau=i\frac{\theta'}{\pi}\,,
\label{ztau}
\ee
according to~\cite{whittaker27}, one has
\be
\left.\frac{\partial\vartheta_1(z|\tau)}{\partial z}\right|_{z=0}
=\frac{1}{(-i\tau)^{3/2}}\left.\frac{\partial\vartheta_1(z|-1/\tau)}{\partial z}\right|_{z=0}\,,
\label{0tau}
\ee
so that
\be
\frac{1}{2q^{1/4}}\left.\frac{\partial\vartheta_1(z,q)}{\partial z}\right|_{z=0}
=\frac{\e^{\theta'/4}}{2}\left(\frac{\pi}{\theta'}\right)^{3/2}
\left.\frac{\partial\vartheta_1(z,\e^{-\pi^2/\theta'})}{\partial z}\right|_{z=0}\,.
\label{0q}
\ee
Accordingly, \eref{T'vtheta'-7} transforms into:
\bea
\fl&\fT'(v,\theta')=\frac{1}{v!(v\!-\!1)!}\nonumber\\
\fl&\ \ \ \ \ \ \ \times\!\prod_{m=0}^{v-1}\left[\frac{\partial}{\partial\theta'}\!+\!m(m\!+\!1)\right]
\left[\left(\frac{\pi}{\theta'}\right)^{3/2}
\e^{-\pi^2/(4\theta')+\theta'/4}\sum_{l=0}^\infty(-1)^l(2l+1)\,\e^{-l(l+1)\pi^2/\theta'}\right]\!.
\label{T'vtheta'-8}
\eea
When $\theta'\ll1$ the leading contribution comes from $l=0$ and the repeated 
application of $\partial/\partial\theta'$, 
so that finally:
\be
\fT'(v,\theta')\simeq\frac{1}{2^{2v}v!(v\!-\!1)!}\left(\frac{\pi}{\theta'}\right)^{2v+3/2}
\e^{-\pi^2/(4\theta')}\,,\qquad\theta'\ll1\,.
\label{T'vtheta'-9}
\ee
The probability density vanishes at $\theta'=0$ with an essential singularity.
The asymptotic behaviours for small and large values of $\theta'$ are shown in figure~\ref{fig-9}.

\section{Discussion}
The probability distribution for the number of particles surviving at a given 
time in the diffusion-coalescence process on the fully-connected lattice (shown in figure~\ref{fig-3}) 
displays Gaussian fluctuations in the scaling limit. 
The mean value $\overline{s_N(t)}$ decays slowly at long time following the $t^{-1}$ behaviour 
obtained in the mean-field approximation~\eref{xt-mf}. After a rapid increase the variance goes 
through a maximum near $t=2$ and then decays asymptotically as $t^{-1}$, too (see figure~\ref{fig-2}.

The probability distribution of the first-passage time through a given number $v$ of surviving particles
behaves quite differently, depending on the value of the particle density $x=v/N$ in the scaling limit.
When $x$ is non-vanishing (i.e., when $v=O(N)$) the mean first-passage time in~\eref{tNxDtN2x}
can be obtained by inverting the mean-field relation~\eref{xt-mf}. The variance in~\eref{tNxDtN2x} 
scales as $N^{-1}$ and increases when $x$ decreases. The fluctuations are Gaussian 
in the scaling limit (see figure~\ref{fig-6}). But the divergence of the variance at $x=0$ signals
the onset of a new scaling behaviour.

Indeed, in the extreme case where $v=O(1)$, the first-passage 
time in figure~\ref{fig-8} displays strong non-Gaussian, non-self-averaging fluctuations. The mean value and 
the standard deviation in~\eref{tNvDtN2v} diverge in the same way, as $N$, in the scaling limit, a characteristic 
of extreme value statistics~\cite{majumdar08,fortin15}. In order to reach a state 
with a few remaining particles the system can follow many alternative paths 
in configuration space, with a wide distribution of arrival times.

It is remarkable that the probability density of the first-passage 
time associated with a number $v$ of surviving particles in~\eref{T'vtheta'-7} 
can be simply obtained by applying a 
product of $v-1$ first-order differential operators on the probability density
obtained for $v=1$:
\bea
\fT'(v,\theta')&=\frac{1}{v!(v\!-\!1)!}\prod_{m=1}^{v-1}\left[\frac{\partial}{\partial\theta'}+m(m+1)\right]
\fT'(1,\theta')\,,\nonumber\\
\fT'(1,\theta')&=\sum_{l=1}^\infty(-1)^{l+1}l(l+1)(2l+1)\,\e^{-l(l+1)\theta'}\,.
\label{op}
\eea
As usual for Gaussian series these probability densities are related to
Jacobi theta functions. They decay exponentially as $\theta'\to\infty$  
and vanish with an essential singularity at $\theta'=0$.

It is interesting to note that similar results have been obtained 
for fluctuating 1D interfaces. In the Edwards-Wilkinson~\cite{edwards82} and 
Kardar-Parisi-Zhang~\cite{kardar86} varieties, the interface configurations in the stationary state
are described by Brownian paths. The probability distribution 
of the squared width, $w^2$, of an interface with periodic boundary conditions, 
involves a universal scaling function of $w^2/\overline{w^2}$ given by a 
Gaussian series~\cite{foltin94}
\be
\Phi(\theta')=\frac{\pi^2}{3}\sum_{l=1}^\infty(-1)^{l+1}l^2\e^{-l^2\theta'}\,,\qquad
\theta'=\frac{\pi^2}{6}\frac{w^2}{\overline{w^2}}\,,
\label{Phi}
\ee
with the same type of asymptotic behaviour as above for large and small values of~$\theta'$. The distribution 
of the maximal relative height, $h_m$, with free or periodic 
boundary conditions, also displays the same type of
asymptotics in the variable $h_m^2/L$, where $L$ is the size of the system~\cite{majumdar04,majumdar05a,majumdar05b,fortin15}.
 
To conclude, among the possible extensions of the present work, let us mention 
the study of density-density correlation functions, $\overline{x(t)x(t')}-\overline{x(t)}\;\overline{x(t')}$, and 
the question of ageing which has been observed for diffusion-coalescence in 1D~\cite{durang10,durang11}. 

\ack
L T gratefully acknowledges helpful discussions with Dragi Karevski and Malte Henkel.

\appendix

\section{Probability distribution \boldmath{$S_N(s,k)$} in the initial state}

In this section we evaluate by induction the components $v_n^{(n)}$ satisfying 
the initial condition $S_N(s,0)=\sum_{n=s}^Nv_s^{(n)}=\delta_{s,N}$. Since $v_s^{(n)}$
defined in~\eref{eigen-3} is proportional to $v_n^{(n)}$, this defines a linear 
system of $N$ equations with $N$ unknowns, $v_n^{(n)}$. One starts with the last 
component $s=N$ which gives the obvious solution $v_N^{(N)}=1$. Then the component 
$v_{N-1}^{(N-1)}$ is solution of $v_{N-1}^{(N-1)}=(N/2)v_{N}^{(N)}=N/2$. The next 
four coefficients are given explicitly by:
\bea
v_{N-2}^{(N-2)}=\frac{1}{4}\frac{(N-1)^2N}{2N-3},\qquad
v_{N-3}^{(N-3)}=\frac{1}{24}\frac{(N-2)(N-1)^2N}{2N-5},\nonumber\\ 
v_{N-4}^{(N-4)}=\frac{1}{96}\frac{(N-3)(N-2)^2(N-1)^2N}{(2N-5)(2N-7)},\nonumber\\
v_{N-5}^{(N-5)}=\frac{1}{960}\frac{(N-4)(N-3)(N-2)^2(N-1)^2N}{(2N-7)(2N-9)}
\eea
A further analysis of these expressions leads to a factorization of $v_n^{(n)}$ as
\be
v_n^{(n)}=\prod_{j=1}^{N-n}\frac{(N-j)(N-j+1)}{j(N+n-j)}={2n-1\choose n}\prod_{j=0}^{n-1}\frac{N-j}{N+j}
\ee
which is the result given in~\eref{vnn} leading to the complete summation formula 
for $S_N(s,k)$ in~\eref{SNsk}. The last expression comes from a rearrangement of the first product.

It will be convenient for later use to introduce the generating function
\be\fl
\CS_N(y,k)=\sum_{s=1}^Nsy^sS_N(s,k)=y\sum_{n=1}^N(2n-1)\prod_{j=0}^{n-1}\frac{N-j}{N+j}
\left[1-\frac{n(n-1)}{N^2}\right]^k\Sigma_n(y)
\label{SNyk-1}
\ee
where, according to~\eref{SNsk}, the last factor is given by
\bea
\Sigma_n(y)&=\sum_{s=1}^n(-1)^{n-s}{n+s-2\choose 2s-2}{2s-2\choose s-1}y^{s-1}\nonumber\\
&=(-1)^{n-1}\sum_{r=0}^{n-1}{r+n-1\choose 2r}{2r\choose r}(-y)^r=P_{n-1}(2y-1)\,,
\label{Sigmany}
\eea
and $P_n(x)$ is the Legendre polynomial. 

The last equality can be demonstrated starting 
with one of the multiple definitions of the Legendre polynomials in terms of derivatives, 
$P_n(x)=(2^nn!)^{-1}d^n[(x^2-1)^n]/dx^n$. 
In particular one has:
\bea
\fl P_n(2y\!-\!1)&\!=\!\frac{1}{n!}\frac{d^n}{dy^n}[y^n(y\!-\!1)^n]
\!=\!\sum_{k=0}^n{n\choose k}^2\!y^{n-k}(y\!-\!1)^{k}
\!=\!\sum_{k=0}^n{n\choose k}^2\!y^{n-k}\sum_{l=0}^k{k\choose l}y^l(-1)^{k-l}\nonumber\\ 
\fl & =\sum_{k=0}^n\sum_{r=n-k}^ny^r{n\choose k}^2{k\choose r-n+k}(-1)^{r-n}\,.
\eea
One can then rearrange the terms in the sum as:
\bea
\fl P_n(2y\!-\!1)&\!=\!\!
\sum_{r=0}^ny^r(\!-\!1)^{r-n}\sum_{k=0}^r\!{n\choose k\!+\!n\!-\!r}^2{k\!+\!n\!-r\!\choose k}
\!\!=\!\!\sum_{r=0}^ny^r(\!-\!1)^{r-n}{n\choose r}\!\sum_{k=0}^r\!{n\choose r\!-\!k}\!{r\choose k}\nonumber\\ 
\fl &=\sum_{r=0}^ny^r(-1)^{r-n}{r+n\choose r}{n\choose r}\,.
\label{Pny}
\eea
The last sum is obtained using the Vandermonde's convolution formula for binomial coefficients. Then 
in~\eref{Sigmany} one can rearrange the product ${r+n-1\choose 2r}{2r\choose r}$ as ${r+n-1\choose r}{n-1\choose r}$,
so that~\eref{Sigmany} can be identified with~\eref{Pny} with $n$ replaced by $n-1$.

Let us now check the initial condition $S_N(s,0)=\delta_{s,N}$ at $k=0$ for which $\CS_N(y,0)=Ny^N$.
According to~\eref{SNyk-1} and~\eref{Sigmany}, one has:
\bea
\frac{\CS_N(y,0)}{y}&=\sum_{s=1}^Nsy^{s-1}S_N(s,0)=\sum_{n=1}^N(2n-1)\prod_{j=0}^{n-1}\frac{N-j}{N+j}P_{n-1}(2y-1)\nonumber\\
&=\sum_{l=0}^{N-1}(2l+1)\prod_{j=0}^l\frac{N-j}{N+j}P_l(2y-1)\,.
\label{SNy0}
\eea
Let rewrite this function as 
\be
\phi_N(x)=\sum_{l=0}^{N-1}(2l+1)\prod_{j=0}^l\frac{N-j}{N+j}P_l(x)=\sum_{l=0}^{N-1}c_lP_l(x)\,,
\label{fNx1}
\ee
the coefficients of the expansion are given by
\be
c_l=(2l+1)\prod_{j=0}^l\frac{N-j}{N+j}=\frac{2l+1}{2}\int_{-1}^1\phi_N(x)P_l(x)dx\,,
\label{cl}
\ee
so that
\be
\frac{1}{2}\int_{-1}^1\phi_N(x)P_l(x)dx=\prod_{j=0}^l\frac{N-j}{N+j}
=\frac{N}{2}\int_{-1}^1\left(\frac{1+x}{2}\right)^{N-1}\!\!\!\!P_l(x)dx\,,
\label{fNx2}
\ee
where the last expression follows from equation 7.127 in~\cite{gradshteyn80}. Finally
\be
\phi_N(x)=N\left(\frac{1+x}{2}\right)^{N-1}\!\!\!\!\!,\qquad \frac{\CS_N(y,0)}{y}=Ny^{N-1}\,,
\label{fNx3}
\ee
and $S_N(s,0)=\delta_{s,N}$ as required.

\section{\boldmath{$\overline{s_N(t)}$} and \boldmath{$\overline{s_N^2(t)}$} when \boldmath{$N\gg1$}}
Let us write:
\be
P_{Nn}=\prod_{j=0}^{n-1}\frac{N-j}{N+j}\,,\qquad E_{Nn}(t)=\left[1-\frac{n(n-1)}{N^2}\right]^{tN}\,.
\label{PE-1}
\ee
For $N\gg1$, the following expansions are obtained
\bea
\fl\ \ \ \ln P_{Nn}&=\sum_{j=1}^{n-1}\left[\ln\left(1-\frac{j}{N}\right)-\ln\left(1+\frac{j}{N}\right)\right]
=-2\sum_{j=1}^{n-1}\left[\frac{j}{N}+\frac{j^3}{3N^3}+{\mathrm O}\left(\frac{j^5}{N^5}\right)\right]\nonumber\\
\fl&=-\frac{n(n-1)}{N}-\frac{[n(n-1)]^2}{6N^3}+{\mathrm O}\left(\frac{n^6}{N^5}\right)\,,\nonumber\\
\fl\ln E_{Nn}(t)&=tN\ln\left[1-\frac{n(n-1)}{N^2}\right]
=-t\left[\frac{n(n-1)}{N}+\frac{[n(n-1)]^2}{2N^3}+{\mathrm O}\left(\frac{n^6}{N^5}\right)\right]\,,
\label{lnPE}
\eea
so that:
\bea
\ \ \ P_{Nn}&=\exp\left[-\frac{n(n-1)}{N}-\frac{[n(n-1)]^2}{6N^3}\right]\left[1+{\mathrm O}\left(\frac{n^6}{N^5}\right)\right]\,,\nonumber\\
E_{Nn}(t)&=\exp\left\{-t\left[\frac{n(n-1)}{N}+\frac{[n(n-1)]^2}{2N^3}\right]\right\}\left[1+{\mathrm O}\left(\frac{n^6}{N^5}\right)\right]\,.
\label{PE-2}
\eea
These expressions are used to evaluate~\eref{sNk} and~\eref{sN2k-2} when $N\gg1$. Then:
\bea
\fl \frac{\overline{s_N(t)}}{N}&\simeq\sum_{n=1}^\infty f(n)\,,\qquad 
\frac{\overline{s_N^2(t)}}{N^2}\simeq\sum_{n=1}^\infty g(n)\,,\nonumber\\
\fl\ \ f(n)&=\frac{2n-1}{N}\exp\left[-\frac{n(n-1)}{N}(t+1)-\frac{[n(n-1)]^2}{6N^3}(3t+1)\right]\,,\nonumber\\
\fl\ \ g(n)&=\frac{(2n-1)}{N}\frac{[1+n(n-1)]}{N}\exp\left[-\frac{n(n-1)}{N}(t+1)-\frac{[n(n-1)]^2}{6N^3}(3t+1)\right]\,.
\label{fg}
\eea
These sums follow from the Euler-Maclaurin formula:
\be\fl
\sum_{n=1}^\infty\rho(n)=\int_0^\infty\!\!\!\rho(n)dn+\frac{1}{2}\left[\rho(\infty)-\rho(0)\right]
+\frac{1}{12}\left[\rho'(\infty)-\rho'(0)\right]+\cdots\,,\quad \rho=f,g\,.
\label{EMf}
\ee
Using the change of variable $u=n(n-1)/N$ together with
$\int_0^\infty du\,u^a\e^{-bu}=\Gamma(a+1)/b^{a+1}$, one obtains:
\bea
\fl\int_0^\infty\!\!\! f(n)dn&=\int_0^\infty\!\!\! du\,\e^{-(t+1)u}-\frac{3t+1}{6N}
\int_0^\infty\!\!\! d u\,u^2\e^{-(t+1)u}+{\mathrm O}\left(N^{-2}\right)\,,\nonumber\\
\fl&=\frac{1}{t+1}-\frac{3t+1}{3N(t+1)^3}+{\mathrm O}\left(N^{-2}\right)\,,\nonumber\\
\fl\int_0^\infty\!\!\!\!g(n)dn&=\int_0^\infty\!\!\!\! du\,u\,\e^{-(t+1)u}+\frac{1}{N}\int_0^\infty\!\!\!\! du\,\e^{-(t+1)u}
-\frac{3t+1}{6N}\int_0^\infty\!\!\!\! d u\,u^3\e^{-(t+1)u}+{\mathrm O}\left(N^{-2}\right)\,,\nonumber\\
\fl&=\frac{1}{(t+1)^2}+\frac{1}{N}\left[\frac{1}{t+1}-\frac{3t+1}{(t+1)^4}\right]+{\mathrm O}\left(N^{-2}\right)\,.
\label{intfg}
\eea
The functions $f(n)$, $g(n)$ and their derivatives vanish exponentially at infinity. With $f(0)=-1/N$ and $f'(0)=2/N$ one has to add a correction 
term $1/3N$ to the integral of $f(n)$ which gives:
\be
\frac{\overline{s_N(t)}}{N}=\frac{1}{t+1}+\frac{1}{3N}\left[1-\frac{3t+1}{(t+1)^3}\right]+{\mathrm O}\left(N^{-2}\right)\,.
\label{sNt}
\ee
The correction to the integral of $g(n)$, of order $N^{-2}$, is negligible so that:
\be
\frac{\overline{s_N^2(t)}}{N^2}=\frac{1}{(t+1)^2}+\frac{1}{N(t+1)}\left[1-\frac{3t+1}{(t+1)^3}\right]+{\mathrm O}\left(N^{-2}\right)\,.
\label{sN2t}
\ee
The variance grows as $N$ and is given by:
\be
\frac{\overline{\Delta s_N^2(t)}}{N}=\frac{1}{3(t+1)}\left[1-\frac{3t+1}{(t+1)^3}\right]+{\mathrm O}\left(N^{-1}\right)\,.
\label{DsN2t}
\ee

\section{Continuum limit of the master equation for \boldmath{$S_N(s,k)$}}

We look for the form of the master equation~\eref{master-1} in the scaling limit. 
Making use of the scaled variables in~\eref{sigmat} with
$\overline{s_N(t)}$ taken from~\eref{sNtDsN2t}, 
the prefactors are rewritten as:
\bea
\fl&1-\frac{s(s-1)}{N^2}=1-\frac{1}{(t+1)^2}-\frac{2\sigma}{N^{1/2}(t+1)}
-\frac{1}{N}\left[\sigma^2-\frac{1}{t+1}\right]+{\mathrm O}\left(N^{-3/2}\right)\,,\nonumber\\
\fl&\ \ \ \ \ \frac{s(s+1)}{N^2}=\frac{1}{(t+1)^2}+\frac{2\sigma}{N^{1/2}(t+1)}
+\frac{1}{N}\left[\sigma^2+\frac{1}{t+1}\right]+{\mathrm O}\left(N^{-3/2}\right)\,.
\label{pref-1}
\eea
The probability density, $\fS(\sigma,t)$, given by the continuum limit 
of $N^{1/2}S_N(s,k)$, depends on $s$ and $k$ through the variables $\sigma(s,k)$ and $t(k)$. A Taylor expansion in $s$ and $k$ 
on the right-hand side of the master equation~\eref{master-1} leads to
\bea
\fl\fS&=\left\{1-\frac{1}{(t+1)^2}-\frac{1}{N^{1/2}}\frac{2\sigma}{t+1}
-\frac{1}{N}\left[\sigma^2-\frac{1}{t+1}\right]\right\}\left[\fS-\frac{\partial\fS}{\partial k}
+\frac{1}{2}\frac{\partial^2\fS}{\partial k^2}\right]\nonumber\\
\fl&\ \ \ \ \ \ \ \ \ \ \ \ \ \ \ \ \ \ +\left\{\frac{1}{(t+1)^2}+\frac{1}{N^{1/2}}\frac{2\sigma}{t+1}
+\frac{1}{N}\left[\sigma^2+\frac{1}{t+1}\right]\right\}\nonumber\\
\fl&\ \ \ \ \ \ \ \ \ \ \ \ \ \ \ \ \ \ \ \ \ \ \ \ \ \ \ \ \ \ \ \ \ \ \ \ \ \times\left[\fS+\frac{\partial\fS}{\partial s}
-\frac{\partial\fS}{\partial k}+\frac{1}{2}\frac{\partial^2\fS}{\partial s^2}
-\frac{\partial^2\fS}{\partial s\,\partial k}+\frac{1}{2}\frac{\partial^2\fS}{\partial k^2}\right]\,,
\label{master-2}
\eea
with partial derivatives given by:
\bea
\frac{\partial\fS}{\partial s}&=\frac{1}{N^{1/2}}\frac{\partial\fS}{\partial\sigma}\,,\qquad
\frac{\partial\fS}{\partial k}=\frac{1}{N^{1/2}(t+1)^2}\frac{\partial\fS}{\partial\sigma}
+\frac{1}{N}\frac{\partial\fS}{\partial t}\,,\nonumber\\
\frac{\partial^2\fS}{\partial s^2}&=\frac{1}{N}\frac{\partial^2\fS}{\partial\sigma^2}\,,\qquad
\frac{\partial^2\fS}{\partial s\,\partial k}=\frac{1}{N(t+1)^2}\frac{\partial^2\fS}{\partial\sigma^2}
+{\mathrm O}\left(N^{-3/2}\right)\nonumber\\
\frac{\partial^2\fS}{\partial k^2}&=\frac{1}{N(t+1)^4}\frac{\partial^2\fS}{\partial\sigma^2}
+{\mathrm O}\left(N^{-3/2}\right)\,.
\label{parder-1}
\eea
Higher derivatives, of order $N^{-3/2}$ or smaller, can be neglected. 

Collecting the coefficients 
of the different powers of $N^{-1/2}$ in~\eref{master-2} the first non-vanishing contributions, 
of order~$N^{-1}$, lead to the partial differential equation~\eref{dSdt}.

\section{Calculation of \boldmath{$\overline{\Delta t_N^2(v)}$} in the scaling limit when $v={\mathrm O}(N)$}

The variance in~\eref{DtN2v} involves the sum
\be
H_{N,2}-H_{v,2}=\sum_{n=v+1}^N\frac{1}{n^2}\,,
\label{HN2Hv2-1}
\ee
which can be evaluated using Euler-Maclaurin formula under the form
\be\fl
\sum_{n=v+1}^N\!\!h(n)\!=\!\int_v^N\!\!\!h(n)dn\!+\!\frac{1}{2}[h(N)\!-\!h(v)]\!+\!\frac{1}{12}[h'(N)\!-\!h'(v)]+\cdots\,,
\qquad h(n)\!=\!\frac{1}{n^2}\,,
\label{EMh}
\ee
leading to:
\be
H_{N,2}-H_{v,2}=\frac{1}{v}-\frac{1}{N}+\frac{1}{2}\left(\frac{1}{N^2}-\frac{1}{v^2}\right)
+\frac{1}{6}\left(\frac{1}{v^3}-\frac{1}{N^3}\right)+\cdots\,.
\label{HN2Hv2-2}
\ee
In the scaling limit, with $v=Nx$, one obtains
\be\fl
2N^2(H_{N,2}-H_{v,2})=-2N\left(1-\frac{1}{x}\right)+1-\frac{1}{x^2}
+\frac{1}{3N}\left(\frac{1}{x^3}-1\right)+{\mathrm O}\left(N^{-2}\right)\,.
\label{HN2Hv2-3}
\ee
Inserting this expression in~\eref{DtN2v}, the leading contribution to the variance is of order $N^{-1}$ and reads:
\be
\overline{\Delta t_N^2(x)}=\frac{1}{N}\left(\frac{2}{3}-\frac{1}{x}+\frac{1}{3x^3}\right)+{\mathrm O}\left(N^{-2}\right)\,.
\label{DtN2x}
\ee

\section{Continuum limit of the master equation for \boldmath{$T_N(v,k)$} when $v={\mathrm O}(N)$}

Using the scaled variables defined in~\eref{thetax} together with $\overline{t_N(x)}$ given by~\eref{tNxDtN2x}, 
the prefactors on the right of the the master equation~\eref{master-3} take the following forms:
\be
1-\frac{v(v+1)}{N^2}=1-x^2-\frac{x}{N}\,,\qquad \frac{v(v+1)}{N^2}=x^2+\frac{x}{N}\,.
\label{pref-2}
\ee
The probability density $\fT(x,\theta)$, which corresponds to $N^{1/2}T_N(v,k)$ in the scaling limit, depends on $v$ and $k$
through the variables $x(v)$ and $\theta(v.k)$. A Taylor expansion in $v$ and $k$ of $\fT$ on the right-hand 
side of the master equation~\eref{master-3} leads to:
\bea
\fl\fT&=\left(1-x^2-\frac{x}{N}\right)\left(\fT-\frac{\partial\fT}{\partial k}
+\frac{1}{2}\frac{\partial^2\fT}{\partial k^2}\right)\nonumber\\
\fl&\ \ \ \ \ \ \ \ \ \ \ \ \ \ \ \ \ \ +\left(x^2+\frac{x}{N}\right)\left(\fT+\frac{\partial\fT}{\partial v}
-\frac{\partial\fT}{\partial k}+\frac{1}{2}\frac{\partial^2\fT}{\partial v^2}
-\frac{\partial^2\fT}{\partial v\,\partial k}+\frac{1}{2}\frac{\partial^2\fT}{\partial k^2}\right)\,.
\label{master-4}
\eea
The partial derivatives are given by:
\bea
\frac{\partial\fT}{\partial v}&=\frac{1}{N^{1/2}x^2}\frac{\partial\fT}{\partial\theta}
+\frac{1}{N}\frac{\partial\fT}{\partial x}\,,\qquad
\frac{\partial\fT}{\partial k}=\frac{1}{N^{1/2}}\frac{\partial\fT}{\partial\theta}\,,\nonumber\\
\frac{\partial^2\fT}{\partial v^2}&=\frac{1}{Nx^4}\frac{\partial^2\fT}{\partial\theta^2}
+{\mathrm O}\left(N^{-3/2}\right)\,,\qquad
\frac{\partial^2\fT}{\partial v\,\partial k}=\frac{1}{Nx^2}\frac{\partial^2\fT}{\partial\theta^2}
+{\mathrm O}\left(N^{-3/2}\right)\nonumber\\
\frac{\partial^2\fT}{\partial k^2}&=\frac{1}{N}\frac{\partial^2\fT}{\partial\theta^2}\,.
\label{parder-2}
\eea
As before, higher derivatives are of order $N^{-3/2}$ or smaller. 

Collecting the coefficients 
of the different powers of $N^{-1/2}$ in~\eref{master-4}, the first non-vanishing contributions are 
of order~$N^{-1}$ and give the partial differential equation~\eref{dTdx}.

\section{Normalization of \boldmath{$\fT'(v,\theta')$}}

Changing the summation index into $l=n-v-1$ in \eref{T'vtheta'-2} gives
\be
\fT'(v,\theta')=\!\frac{1}{v!(v\!-\!1)!}\sum_{l=0}^\infty(-1)^l(2l\!+\!2v\!+\!1)
\frac{(l\!+\!2v)!}{l!}\,\e^{-(l+v)(l+v+1)\theta'}\,,
\label{T'vtheta'-3}
\ee
so that:
\bea
\fl\CN\!=\!\!\!&\int_0^\infty \fT'(v,\theta')\,d\theta'=\frac{(2v)!}{v!(v\!-\!1)!}\sum_{l=0}^\infty(-1)^l
\frac{2l\!+\!2v\!+\!1}{(l+v)(l+v+1)}{l+2v\choose l}\nonumber\\
\fl&\ \ \ \ \ \ \ \ \ \ \ \ \ \ \ \ \ \ \ \ \ \ =\frac{(2v)!}{v!(v\!-\!1)!}\sum_{l=0}^\infty(-1)^l\left(\frac{1}{l+v}+\frac{1}{l+v+1}\right){l+2v\choose l}\nonumber\\
\fl&\ \ \ \ \ \ \ \ \ \ \ \ \ \ \ \ \ \ \ \ \ \ =\frac{(2v)!}{v!(v\!-\!1)!}\Bigg[\underbrace{\sum_{l=0}^\infty\frac{(-1)^l}{l+v}{l+2v\choose l}}_{S_0(1)}+
\underbrace{\sum_{l=0}^\infty\frac{(-1)^l}{l+v+1}{l+2v\choose l}}_{S_1(1)}\Bigg]\,.
\label{norm-1}
\eea
Here $S_0(1)$ and $S_1(1)$ are the values of the function
\be
S_n(y)=\sum_{l=0}^\infty\frac{(-1)^l}{l+v+n}{l+2v\choose l}y^{l+v+n}
\label{Sny}
\ee
when $y\to1_-$.
Its derivative is given by:
\be
\frac{dS_n}{dy}=\sum_{l=0}^\infty(-1)^l{l+2v\choose l}y^{l+v+n-1}=\frac{y^{v+n-1}}{(1+y)^{2v+1}}\,,\qquad |y|<1\,.
\label{S'ny}
\ee
Since $v>0$ one has $S_n(0)=0$ and
\be\fl
S_0(1)+S_1(1)=\lim_{\epsilon\to0}\int_0^{1-\epsilon}\left[\frac{y^{v-1}}{(1+y)^{2v+1}}+\frac{y^v}{(1+y)^{2v+1}}\right]dy=\int_0^1\frac{y^{v-1}}{(1+y)^{2v}}dy\,.
\label{S0+S1-1}
\ee
With the change of variables $y=z/(1-z)$ one obtains
\be
S_0(1)+S_1(1)=\int_0^{1/2}z^{v-1}(1-z)^{v-1}dz=\frac{B(v,v)}{2}=\frac{v!(v\!-\!1)!}{(2v)!}\,,
\label{S0+S1-2}
\ee
where $B(v,v)$ is the Euler beta function. Thus, according to~\eref{norm-1}, $\CN=1$,
as required.

Actually $\CN$ in \eref{norm-1} is an alternate divergent series which can be summed using the original method of 
Euler~\cite{euler1754} which goes as follows (see~\cite{hardy1949,sandifer06}). Given the alternate series $\CN=a-b+c-d+e-\cdots$, build the sequences of
successive finite differences
\bea
\Delta^1&=b-a,c-b,d-c,e-d,\ldots\quad&\alpha=b-a\,,\nonumber\\
\Delta^2&=c-2b+a,d-2c+b,e-2d+c,\ldots\quad&\beta=c-2b+a\,,\nonumber\\
\Delta^3&=d-3c+3b-a,e-3d+3c-b,\ldots\quad&\gamma=d-3c+3b-a\,,\nonumber\\
&\ \ \vdots &\ \ \ \vdots
\label{dif-1}
\eea
until eventually the differences vanish. Then the sum is given by:
\be
\CN=\frac{a}{2}-\frac{\alpha}{4}+\frac{\beta}{8}-\frac{\gamma}{16}+\cdots\,.
\label{sum}
\ee
For example, when $v=2$ 
\bea
\CN&=10-35+81-154+260-405+\cdots\ \ \ &a=10\,,\nonumber\\
\Delta^1&=25,46,73,106,145,\ldots\quad &\alpha=25\,,\nonumber\\
\Delta^2&=21,27,33,39,\ldots\quad &\beta=21\,,\nonumber\\
\Delta^3&=6,6,6,\ldots\quad &\gamma=6\,,\nonumber\\
\Delta^4&=0,0,\ldots\quad &\delta=0\,,
\label{dif-3}
\eea
and $\CN=10/2-25/4+21/8-6/16=1$.

\section*{References}
\bibliographystyle{iopart-num}
\bibliography{biblio_readif}

\end{document}